\documentclass[preprint]{ptephy_v1}
\usepackage[ascii]{inputenc}
\usepackage[T3,T1]{fontenc}
\usepackage[english]{babel}
\usepackage[noenc]{tipa}
\usepackage{tipx}
\usepackage[geometry,weather,misc,clock]{ifsym}
\usepackage{pifont}
\usepackage{eurosym}
\usepackage{amsmath}
\usepackage{wasysym}
\usepackage{amssymb,amsfonts,textcomp}
\usepackage{array}
\usepackage{supertabular}
\usepackage{hhline}
\usepackage{graphicx}
\makeatletter
\newcommand\arraybslash{\let\\\@arraycr}
\makeatother
\renewcommand\arraystretch{1.3}
\providecommand{\keywords}[1]{\textbf{\textit{Index terms---}} #1}

\date{2017-01-10}
\begin{document}

\title{Theoretical studies on mechanical and electronic properties of $s$-triazine sheet}


\author{Yusuf Zuntu Abdullahi}
\affil{School of Physics, Universiti Sains Malaysia, 11800 Penang, Malaysia}
\affil{Department of Physics, Faculty of Science, Kaduna State University, P.M.B. 2339 Kaduna State, Nigeria 
 \email{yusufzuntu@gmail.com}}

\author{Tiem Leong Yoon}
\affil{School of Physics, Universiti Sains Malaysia, 11800 Penang, Malaysia
\email{tlyoon@usm.my}}

\author{Mohd Mahadi Halim}
\affil{School of Physics, Universiti Sains Malaysia, 11800 Penang, Malaysia}

\author{Md. Roslan Hashim}
\affil{Institute of Nano-Optoelectronics Research and Technology Laboratory, Universiti Sains Malaysia, 11900 Penang, Malaysia}

\author{Thong Leng Lim}
\affil{Faculty of Engineering and Technology, Multimedia University, Jalan Ayer Keroh Lama, 75450 Melaka, Malaysia}

\begin{abstract}

Mechanical and electronic properties of $s$-triazine are studied using first-principles calculations based on density functional theory. The in-plane stiffness and bulk modulus for $s$-triazine sheet are found to be less than that of heptazine. The reduction can be related to the nature of the covalent bonds connecting the adjacent sheets and the number of atoms per unit cell. The Poisson's ratio of $s$-triazine is half the value to that of graphene. Additionally, the calculated values of the two critical strains (elastic and yielding points) of $s$-triazine sheet are in the same order of magnitude to that for heptazine which was calculated using MD simulations in the literature. It is also demonstrated that $s$-triazine sheet can withstand larger tension in the plastic region. These results established a stable mechanical property for $s$-triazine sheet. We found a linear relationship of bandgap as a function of bi-axial tensile strain within the harmonic elastic region. The reduced steric repulse of the lone pairs ($\mathrm{p}_x$-, $\mathrm {p}_y$-) causes the $\mathrm {p}_z$-like orbital to shift to high energy, and consequently an increase in the bandgap. We find no electronic properties modulation of the $s$-triazine sheet under electric field up to a peak value of 10 V/nm. Such noble properties may be useful in future nanomaterial applications.

\end{abstract}


\maketitle

\section{Introduction}
$s$-triazine is one of the newly found allotropes of two-dimensional graphitic carbon nitride sheet with a chemical formula of ${\mathrm C}_6{\mathrm N}_6$ per unit cell. It has been found to be dynamically and structurally stable by phonon calculation based on density functional theory (DFT) and molecular dynamic (MD) simulations respectively at ambient temperature [1]. The semiconducting property of $s$-triazine ($g$-${\mathrm C}_6{\mathrm N}_6$) is comparable to that of heptazine ($g$-${\mathrm C}_6{\mathrm N}_7$) in the absence of spin orbit coupling (SOC). Inclusion of SOC in DFT calculations of $g$-${\mathrm C}_6{\mathrm N}_6$ produces a topological invariant of $Z=1$ [1]. This qualifies $s$-triazine material for quantum spin hall effect (QSHE) applications at a temperature less than 95 K [2, 3].

In the last few years, much attention has been focused on theoretical understanding of geometric and electronic properties of pure and doped graphitic carbon nitride (CN) 2D allotropes [4-13]. However, only few computational studies on modulations of mechanical and electronic properties under applied tensile strain of graphitic CN allotropes are available. Very recently a full atomistic reactive MD simulation has shown that $g$--${\mathrm C}_6{\mathrm N}_7$ is mechanically stable at a maximum of 600K and has been found to exhibit less fractures under larger tensions. They also observe that the fracture pattern under larger tensions depends on the chemical bonds, density values, topologies and stretching directions [14].

In our recent DFT calculations, we found that heptazine material can withstand large tension in the plastic region [15]. For in-plane stiffness (Young modulus), our previous result for heptazine is significantly higher than that reported in [14]. Our previous work also predicts increased bandgap caused by symmetric deformations as a result of applied bi-axial tensile strain within the elastic region [15]. It has also been reported that electronic and magnetic properties of graphitic triazine based CN ($g$-$\mathrm{C}_4\mathrm{N}_3)$ are modulated as a result of asymmetric deformations induced by uniaxial tensile strain [16]. Therefore, it will be interesting to investigate mechanical properties and strain-induced responses on electronic properties based on DFT calculations of $s$-triazine sheet as one of the allotropes of graphitic 2D CN sheet.

\section{Calculation method}
First-principles calculations have been carried out based on DFT [17] using the generalized gradient approximation (GGA) of Perdew-Burkew-Enzerhof (PBE) [18] functional as implemented in the QUANTUM ESPRESSO code [19]. To treat electron-ion core interaction ultrasoft pseudopotentials have been adopted for C, N atoms [20]. The plane wave cutoff energy up to 500 eV has been used. 10{\texttimes}10{\texttimes}1 Monkhorst-Pack special k-point set has been adopted for 2D Brillouin zone (BZ) integration [21]. Density of states have been computed with denser set of 18{\texttimes}18{\texttimes}1 k-points. To avoid interaction between the $s$-triazine sheet and its neighboring image in a perpendicular direction, a vacuum of about 16 {\AA} was used. We consider a  $1\times 1$ pristine unit cell of $s$-triazine sheet containing 6 carbon and 6 nitrogen atoms. We have used our computed equilibrium lattice parameter for $s$-triazine of 7.13 {\AA} which agrees well with the previously reported theoretical works [1]. The criteria of convergence for forces on each atom within Broyden--Fletcher--Goldfarb--Shanno (BFGS) and total energy were 0.002 eV/{\AA} and 10$-$4 eV respectively.

\section{Results and discussion}
We first examine the optimized structural properties of the pristine $s$-triazine sheet. As shown in Figure. 1, each carbon atom is three-coordinated with two nitrogen atoms and a C atom linking to a separate  $s${}-triazine. Hence, the C atom in the  $s${}-triazine has a $\mathrm {sp}^2${}-like hybridized structure. The N atoms are two-coordinated with $\mathrm{sp}^3${}-like hybridized structure. The optimized C-C bond length linking adjacent $s$-triazine unit is found to be 1.51 {\AA}, which agrees well with the previous work [1] and tallies with the bond length of diamond of 1.52 {\AA}. The angles formed by N-C-N (within the $s$-triazine rings) and N-C-C (within the cavity) are measured to be  $125.68{}^{\circ}$ and  $117.17{}^{\circ}$ respectively, confirming a  $\sigma ${}-like orbitals $\mathrm {sp}^2$ hybridized structure. The length of the C-N bond is slightly lower than the bond length of diamond, 1.34 {\AA}; whereas the C-N-C angle at  $114.33{}^{\circ}$ is slightly larger than  $109.5{}^{\circ}$ found in diamond structure. These results are consistent with previous work [1] and suggestive of $\mathrm {sp}^3${}-like hybridized structure of N atom.

We further validate the bonding nature of $s$-triazine sheet by plotting the charge density distribution which is shown in the right side of Figure.1. The plot shows a denser contour around the N atoms, displaying superior electronegativity of N as compared to C atom. The bond length weakening of C-C and stiffness of C-N are evidenced by the charge density values of 0.4632 and 0.579 (in a.u.) respectively. These results confirm the reliability of our computational parameters used.

Next we examine the mechanical properties/stabilities of the $s$-triazine sheet. The mechanical properties were carefully quantified using strain energy as a function of uni- and bi-axial tensile strains within a linear region ranging from -2\% to 2\% in a step of 0.005 as shown in Figures.2a, 2b. The in-plane stiffness (Young modulus) $Y$ and Poisson's ratio $\nu$ were then evaluated via the following equations:
\begin{eqnarray}
C_{11}&=&{1 \over A_0}\left[ {\partial^2 E \over \partial s^2}\right]_{s=0} \ \ \ (\textrm{uniaxial}) \nonumber \\
2(C_{11}+C_{12})&=&{1 \over A_0}\left[ {\partial^2 E \over \partial s^2}\right]_{s=0} \ \ \ (\textrm{bi-axial})
\end{eqnarray}
In-plane stiffness and Poisson's ratio:
\begin{eqnarray}
Y&=&C_{11}(1 - \nu^2), \nonumber \\
\nu &=& {C_{12} \over C_{11}}
\end{eqnarray}
where $C_{11}, C_{12}$ denote the elastic constant, $A_0, E$, and $s$ are the equilibrium unit-cell area, strain energy and applied tensile strain. The calculated in-plane stiffness and Poisson's ratio are  $1335.5 \ \mathrm{GPa}{\cdot}\text{{\AA}}$ (= 133.55 N/m) and 0.08 respectively. The in-plane stiffness for $s$-triazine is approximately lower than that for heptazine [15] by 78.9 N/m and graphyne [22] by 57.41 N/m respectively. The reduced value of in-plane stiffness in comparison with heptazine can be related to the structural futures of the sheet in terms of bonds linking the adjacent $s$-triazine sheet, atomic configuration and the number of atoms per unit cell. On the hand, the estimated Poisson's ratio is exactly half the value of graphene reported in [23]. The bulk modulus is estimated from second derivative of the total strain energy as a function of area of the s-triazine sheet defined by Eq. (3),

\begin{equation}
G=A\times \left.\left(\frac{{\partial}^2E}{{\partial}A^2}\right)\right|_{A_m}
\end{equation}
where $A$, $E$ and $A_m$ are the area of the s-triazine unit cell, the total strain energy and the unit cell area of the equilibrium structure respectively. Figure. 2c shows dependency of strain energy on area of $s$-triazine sheet. The estimated value for the bulk modulus is 82.8 N/m which is also lower than the value for heptazine sheet [15]. Overall, all the findings of mechanical properties values of $s$-triazine sheet are within the order of magnitude for heptazine sheet, clearly indicating that $s$-triazine is mechanically stable. \ \ 

To examine the effect of mechanical strain on the structural stabilities of $s$-triazine sheet, we applied both uni- and bi-axial tensile strain s on the $s$-triazine sheet by gradually increasing the lattice constant which is expressed as $s=\Delta a / a_0$,
where   
$a_0$ is the equilibrium lattice constant of the $s${}-triazine sheet. In Figure. 3, we show strain energy ($E$) and first derivative ($\frac{{\partial}E}{{\partial}s}$) of the strain energy of  $s${}-triazine sheet as functions of the uni-axial strain (in Figure 3a, 3b) and bi-axial strain (in Figure. 3c, 3d) respectively. It is observe in Figure. 3 that an increase in tensile strain (both uni- and bi-axial deformations) causes a corresponding change in the strain energy until it attains a constant value. Consequently, two critical points can be deduced for both uni- and bi-axial from Figures. 3b, 3d, (from the  $\frac{{\partial}E}{{\partial}s} \ \mathrm{vs}. \ s$ curves). First critical point (depicted by circled star) is the point below which  $\frac{{\partial}E}{{\partial}s}$ grows linearly with tensile strain  $s$ (see Figures. 3b, 3d). We denote the point of proportionality at which the strain energy increases linearly with increasing tensile strain by  $s_{\mathrm{m1}}$ (see Figures. 3a, 3c). Hence,  $0{\leq}s{\leq}s_{\mathrm{m1}}$ is the linear elastic region. On the other hand, it can be observed that the  $\frac{{\partial}E}{{\partial}s} \ \mathrm {vs}. \ s$ curves gradually move away from a linear relation when the tensile strain is increased above $s_{\mathrm m1}$ point below $s_{\mathrm m2}$. Consequently, the slope of \  $\frac{{\partial}E}{{\partial}s}$ also gradually decreases until $s$ reaches $s_{\mathrm m2}$ point. The point  $s=s_{\mathrm{m2}}$,  $\left.\frac{{\partial}E}{{\partial}s}\right|_{s_2}=0$ where the stress reaches its peak value is called the yielding point.  $0{\leq}s{\leq}s_{\mathrm{m2}}$ is the elastic region, in which $s$-triazine can regains its original shape and structure upon release of applied strain. One can obviously observe in Figure. 3b, 3d that beyond $s_{\mathrm m2}$, the  $\frac{{\partial}E}{{\partial}s} \ \mathrm{vs.} \ s$ curve abruptly drops. We can infer that a permanent deformation of the system has occurred at  $s{\geq}s_{\mathrm{m2}}$. This region is called plastic region. 

We also verify the results of in-plane stiffness and Poisson's ratio obtained in previous paragraphs using harmonic potential of the form  $E=\frac 1 2ks^2$, where  $k=\left.\frac{{\partial}^2E}{{\partial}s^2}\right|_{E_{\mathrm{min}}}$,  $E_{\mathrm{min}}=\mathrm{min}\left[E\right]$. It is numerically evaluated by fitting the strain energy curves within the linear elastic region (see the Figures. 3a, 3c). The computed results are listed in Table 2. Accordingly,  $s${}-triazine sheet can exhibit stable mechanical properties and withstand longer tensions in the plastic region as was observed in heptazine sheet [15]. 

In comparison, the two critical strains  $s_{\mathrm{m1}}$,  $s_{\mathrm{m2}}$ calculated in this work show a significant improvement to that for heptazine calculated by us recently [15]. The main difference in critical strains between the $s$-triazine sheet in this work and that of heptazine sheet in [15] lies in the geometric configurations. We find that the adjacent $s$-triazine units are connected via a C-C bond which bond length is the same as that in diamond, whereas adjacent heptazine units are linked via C-N bonds. Thus the C-C bonds are relatively stronger than the C-N bonds. It also interesting to note that despite the discrepancy in critical strains, both sheets ($s$-triazine and heptazine) can withstand larger tension in their plastic regions. 

While we find some differences in values of the calculated critical strains and the mechanical properties of $s$-triazine in this work and heptazine reported in [15], de Sousa et al. recent work reported some numerical results which are not much different from our present work [14]. In their molecular-dynamics (MD) calculations, ReaxFF forcefield is employed to model the interactions between the heptazine atoms. Opposed to being isotropic, as is assumed in this work, they consider anisotropic heptazine sheet. They calculated critical strains at various temperatures whereas our DFT calculations assume that temperature is implicitly set at 0 K. It is found that the critical strain reported in their work at 10 K agrees with our present result, i.e., it is 0.149 along the (armchair $x${}-axis), while that along the zigzag ($y${}-axis), 0.178. Their computed values are at the order or magnitude as our calculated critical strains $s_{\mathrm{m2}}$. We expect that the value should agree with their estimations since their critical strains are set at the point where fracture starts. Of another special interest is the Young modulus reported at 10 K, which is 1397 $\mathrm{GPa}{\cdot}\text{{\AA}}$ in the  $x${}-axis and  1356 $\mathrm{GPa}{\cdot}\text{{\AA}}$ in the  $y${}-axis, which tally well with our computed in-plane stiffness of  1335.5 $\mathrm{GPa}{\cdot}\text{{\AA}}$. Despite the differences in computational procedure between MD and DFT calculations, we observe no strong discrepancies in the obtained numerical results between our work and that reported in [14].

To obtain an insight into the bandgap tuning of $s$-triazine sheet under symmetric deformation, we first analyze the electronic band structure and the corresponding density of state (DOS) of strain-free  $s${}-triazine. Figures. 5a, c confirms non-magnetic and semiconducting properties of of $s$-triazine sheet with a direct bandgap of approx. 1.53 eV [1]. Both the valence band maximum (VBM) and the conduction band minimum (CBM) are situated at the K point in the Brillouin zone. From the illustrated Figures. 5e, 5g of projected density of state (PDOS) for C and N of unstrained  $s${}-triazine sheet, it can be seen that the top valence band is dominated by  $\sigma ${}-like orbital states (px, py) including s orbital, whereas the bottom of the conduction band in the vicinity of the Fermi level has a dominant feature of $\mathrm {p}_z$-like orbitals from both C and N. 

When symmetric deformation (bi-axial tensile strain) is applied, bandgap is increased (see Figure. 4). This is obviously shown in the band structure and DOS Figures. 5b, d. The increase in bandgap can be explained as follows. From the PDOS Figures. 5f, h of strained $s$-triazine sheet (3\% bi-axial tensile strain), we observe a uniform shift of $\mathrm {p}_z$-like states towards higher energy as the bi-axial tensile strain is applied, resulting in an increased bandgap. As previously reported in our recent work, the symmetric deformations make the lone pairs (contributed by px, py) to become disoriented, thereby causing $\mathrm {p}_z$-like orbital to shift towards higher energy as a result of reduced steric repulsions of the px, py lone pairs. We also observe no change in structural (see Figure.1 side view of the $s$-triazine sheet) and electronic (see Figure. 5i) properties at a maximal of 10 V/nm perpendicular electric field amplitude compared to the results at zero electric field (see Figure. 1 and Figure. 5a).

\section{Summary}
In this study we use DFT calculations to examine the mechanical and electronic properties of $s$-triazine sheet. We have shown that the in-plane stiffness and bulk modulus for $s$-triazine sheet have lower values than that of heptazine. The lower value of in-plane stiffness and bulk modulus in comparison with heptazine sheet can be linked to the inherent structural futures of the $s$-triazine sheet in terms of covalent bonds linking the adjacent sheets and the number of atoms per unit cell. The Poisson's ratio of $s$-triazine is half the value to that of graphene. Interestingly, our computed values of the two critical strains (elastic and yielding points) of $s$-triazine sheet are in the same order of magnitude to that of breaking point for heptazine material which was reported using MD simulations. We have also shown that the $s$-triazine sheet can withstand larger tension in the plastic region. There is a linear relation between bandgap and the symmetric deformation as a result of applied bi-axial tensile strain. Reduction in the steric repulsion of the lone pairs causes the $\mathrm {p}_z$-like orbitals to shift towards higher energy, rendering an increase in the bandgap. We find no modulations in the electronic properties of the sheet under electric field up to a maximal value of 10 V/nm.

\section*{Acknowledgments}

T. L. Yoon wishes to acknowledge the support of Universiti Sains Malaysia RU grant (No. 1001/PFIZIK/811240). Figure showing atomic model and 2D charge-density plots are generated using the XCRYSDEN program Ref. [24]. We gladfully acknowledge Dr. Chan Huah Yong from the School of Computer Science, USM, for providing us computing resources to carry out part of the calculations done in this paper.

\section*{References}

[1]\ \ A. Wang, X. Zhang, M. Zhao, Topological insulator states in a honeycomb lattice of s-triazines, Nanoscale., 6 (2014), pp. 11157-11162.

[2]\ \ Y. Yao, F. Ye, X.-L. Qi, S.-C. Zhang, Z. Fang, Spin-orbit gap of graphene: First

principles calculations, Phys. Rev. B: Condens. Matter Mater. Phys., \ 75 (2007), pp. 041401.

[3]\ \ H. Min, J. E. Hill, N. A. Sinitsyn, B. R. Sahu, L. Kleinman, A. H. MacDonald, Intrinsic and Rashba spin-orbit interactions in graphene sheets, Phys. Rev. B: Condens. Matter Mater. Phys., 74 (2006), pp. 165310.

[4]\ \ A. Du, S. Sanvito, S.C. Smith, First-principles prediction of metal-free magnetism and intrinsic half-metallicity in graphitic carbon nitride, Phys. Rev. Lett., 108 (2012), pp. 197207.

[5]\ \ X. Li, S. Zhang, Q. Wang, Stability and physical properties of a tri-ring based porous gC 4 N 3 sheet, Phys. Chem. Chem. Phys., 15 (2013), pp. 7142-7146.

[6]\ \ P. Niu, G. Liu, H.-M. Cheng, Nitrogen vacancy-promoted photocatalytic activity of graphitic carbon nitride, J. Phys. Chem. C., 116 (2012), pp. 11013-11018.

[7]\ \ P. Niu, L. Zhang, G. Liu, H.M. Cheng, Cheng, Graphene{}-Like Carbon Nitride Nanosheets for Improved Photocatalytic Activities, Adv. Funct. Mater., 22 (2012), pp. 4763-4770.

[8]\ \ D.M. Teter, R.J. Hemley, Low-compressibility carbon nitrides, Science 271 (1996), pp. 53-55.

[9]\ \ Y.Z. Abdullahi, T.L. Yoon, M.M. Halim, M.R. Hashim, M.Z.M. Jafri, L.T. Leng, Geometric and electric properties of graphitic carbon nitride sheet with embedded single manganese atom under bi-axial tensile strain, Curr. Appl. Phys., 16 (2016), pp. 809-815.

[10]\ \ D. Ma, Q. Wang, X. Yan, X. Zhang, C. He, D. Zhou, Y. Tang, Z. Lu, Z. Yang, 3d transition metal embedded C 2 N monolayers as promising single-atom catalysts: A first-principles study Carbon, 105 (2016), pp. 463-473.

[11]\ \ D. Ghosh, G. Periyasamy, B. Pandey, S.K. Pati, Computational studies on magnetism and the optical properties of transition metal embedded graphitic carbon nitride sheets, J. Mater. Chem. C., 2 (2014), pp. 7943-7951.

[12]\ \ S. Zhang, R. Chi, C. Li, Y. Jia, Structural, electronic and magnetic properties of 3d transition metals embedded graphene-like carbon nitride sheet: A DFT+ U study, Phys. Lett. A., 380 (2016), pp. 1373-1377.

[13]\ \ I. Choudhuri, P. Garg, B. Pathak, TM@ gt-C 3 N 3 monolayers: high-temperature ferromagnetism and high anisotropy, J. Mater. Chem. C., 4 (2016), pp. 8253-8262.

[14]\ \ J. M. de Sousa, T. Botari, E. Perim, R. A. Bizao, D. S. alvGao, Mechanical and structural properties of graphene-like carbon nitride sheets, RSC Adv., 6 (2016) pp. 76915-76921.

[15]\ \ Y. Z. Abdullahi, T. L. Yoon, M. M. Halim, M. R. Hashim, T. L. Lim, Mechanical and electronic properties of graphitic carbon nitride sheet: First-principles calculations, Solid State Commun., 248 (2016), pp. 144-150.

[16]\ \ L. Liu, X. Wu, X. Liu, P.K. Chu, Electronic structure and magnetism in g-C4N3 controlled by strain engineering, Appl. Phys. Lett., 106 (2015), pp. 132406.

[17]\ \ P. Hohenberg, W. Kohn, Inhomogeneous electron gas, Phys Rev., 136 (1964), pp. B864.

[18]\ \ J.P. Perdew, K. Burke, M. Ernzerhof, Generalized gradient approximation made simple, Phys. Rev. Lett., 77 (1996), pp. 3865.

[19]\ \ P. Giannozzi, S. Baroni, N. Bonini, M. Calandra, R. Car, C. Cavazzoni, D. Ceresoli, G.L. Chiarotti, M. Cococcioni, I. Dabo, QUANTUM ESPRESSO: a modular and open-source software project for quantum simulations of materials, J. Phys. Condens. Matter., 21 (2009), pp. 395502.

[20]\ \ D. Vanderbilt, Soft self-consistent pseudopotentials in a generalized eigenvalue formalism, Phys. Rev. B., 41 (1990), pp. 7892.

[21]\ \ H.J. Monkhorst, J.D. Pack, Special points for Brillouin-zone integrations, Phys. Rev. B., 13 (1976), pp. 5188.

[22]\ \ M. Asadpour, S. Malakpour, M. Faghihnasiri, B. Taghipour, Mechanical properties of two-dimensional graphyne sheet, analogous system of BN sheet and graphyne-like BN sheet, Solid State Commun., 212 (2015), pp. 46-52.

[23]\ \ E. Cadelano, P.L. Palla, S. Giordano, L. Colombo, Elastic properties of hydrogenated graphene, Phys. Rev. B., 82 (2010), pp. 235414.

[24]\ \ A. Kokalj, Computer graphics and graphical user interfaces as tools in simulations of matter at the atomic scale, Computer. Mater. Sci., 28 (2003), pp. 155-168.

\begin{table}
\label{table1}
\begin{center}
\tablefirsthead{}
\tablehead{}
\tabletail{}
\tablelasttail{}
\begin{supertabular}{m{0.6205598in}m{0.83445984in}m{1.0462599in}m{0.97755986in}m{0.82615983in}}
\hline
{\centering Strain\par}

\centering (\%) &
{\centering Area\par}

\centering ({\AA}2) &
{\centering Total energy\par}

{\centering Biaxial\par}

\centering (Ry) &
{\centering Total energy\par}

{\centering Uniaxial\par}

\centering (Ry) &
{\centering Lattice parameter\par}

{\centering Biaxial\par}

\centering\arraybslash ({\AA})\\\hline
\centering {}-0.02 &
\centering 42.32 &
\centering {}-187.44415 &
\centering {}-210.08377 &
\centering\arraybslash 6.99/12.11\\
\centering {}-0.015 &
\centering 42.76 &
\centering {}-187.45033 &
\centering {}-210.09083 &
\centering\arraybslash 7.03/12.17\\
\centering {}-0.01 &
\centering 43.19 &
\centering {}-187.45457 &
\centering {}-210.09547 &
\centering\arraybslash 7.06/12.23\\
\centering {}-0.005 &
\centering 43.63 &
\centering {}-187.456953 &
\centering {}-210.09774 &
\centering\arraybslash 7.10/12.29\\
\centering 0 &
\centering 44.07 &
\centering {}-187.457620 &
\centering {}-210.09773 &
\centering\arraybslash 7.13/12.36\\
\centering 0.005 &
\centering 44.51 &
\ \ \ {}-187.456549 &
\centering {}-210.09557 &
\centering\arraybslash 7.17/12.42\\
\centering 0.01 &
\centering 44.95 &
{\centering {}-187.453911\par}

~
 &
\centering {}-210.09152 &
\centering\arraybslash 7.21/12.48\\
\centering 0.015 &
\centering 45.40 &
\centering {}-187.4497165 &
\centering {}-210.08562 &
\centering\arraybslash 7.24/12.54\\
\centering 0.02 &
\centering 45.85 &
\centering {}-187.444124 &
\centering {}-210.07790 &
\centering\arraybslash 7.28/12.60\\\hline
\end{supertabular}
\caption{The optimized lattice parameters and total energy of $s$-triazine sheet for bulk modulus estimations.}
\end{center}
\end{table}


\begin{table}
\label{table2}
\tablefirsthead{}
\tablehead{}
\tabletail{}
\tablelasttail{}
\begin{center}
\begin{supertabular}{|m{1.9795599in}|m{1.1087599in}|m{0.9219598in}|}
\hline
~
&
\centering Uni-axial &
\centering\arraybslash Bi-axial\\\hline
Harmonic constant,  $k$ (N/m) &
\centering 134.41 &
\centering\arraybslash 290.10\\\hline
\centering Proportionality limit,  $s_{\mathrm{m1}}$ &
\centering 0.086 &
\centering\arraybslash 0.065\\\hline
\centering Yielding strain,  $s_{\mathrm{m2}}$ &
\centering 0.107 &
\centering\arraybslash 0.112\\\hline
\centering  $\nu $, Poisson's ratio &
\multicolumn{2}{m{2.1094599in}|}{\centering 0.08}\\\hline
\centering  $E$, in-plane stiffness  &
\multicolumn{2}{m{2.1094599in}|}{\centering 1335.5 \ $\mathrm{GPa}{\cdot}\text{{\AA}}$}\\\hline
\centering  $G$, Bulk modulus  &
\multicolumn{2}{m{2.1094599in}|}{\centering 828 \  $\mathrm{GPa}{\cdot}\text{{\AA}}$}\\\hline
\end{supertabular}
\caption{Numerically calculated mechanical properties obtained from the strain energy curves in Figure 3.}
\end{center}
\end{table}

\begin{figure}
\label{fig1}
\includegraphics[width=6.4965in,height=1.8272in]{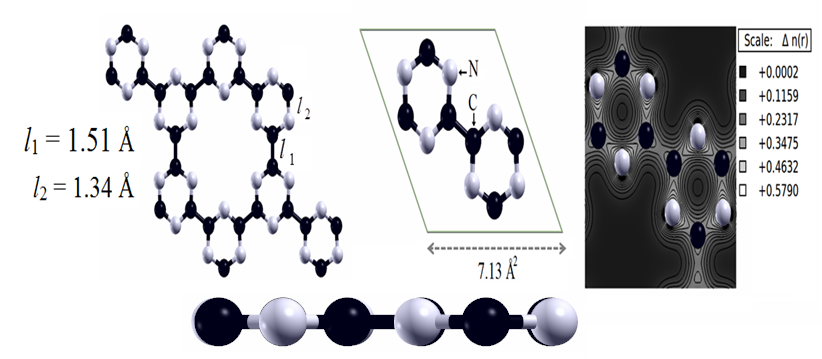}
\caption{Middle: Top view of the  $1\times 1$ $s$-triazine unit cell. The atomic symbols and the computed lattice constant of 7.14 {\AA} are well labeled. Right: Top view of the charge-density distributions of the nearest neighboring C and N atoms of a unstrained $s$-triazine, with contours indicating charge accumulation and color ranges in a.u. carbon atoms are in black, whereas nitrogen atoms in white color. Bottom: Side view of the $s$-triazine sheet under electric field of magnitude 10.0 V/nm}
\end{figure}

\begin{figure}
\begin{center}
\label{fig2}
(a)  
\includegraphics[width=2.9063in,height=2.302in]{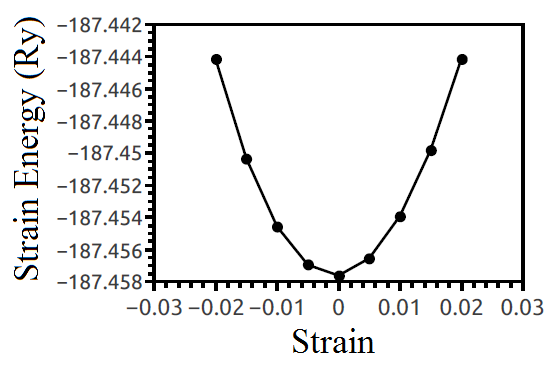}
(b)  \includegraphics[width=3.1043in,height=2.3126in]{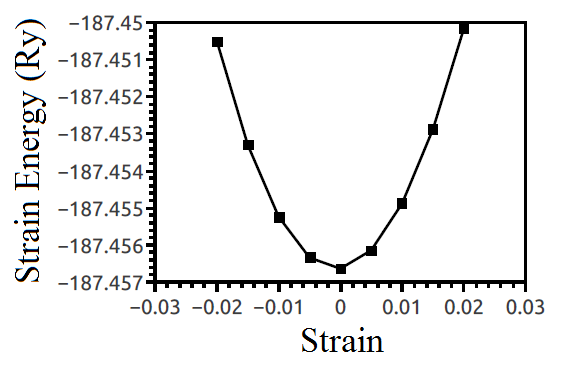}
(c) \includegraphics[width=3.0102in,height=2.3437in]{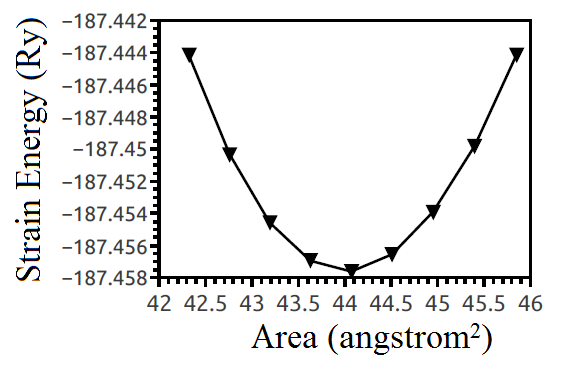}
\caption{
(a). The dependence of strain energy (Ry) on bi-axial tensile strain of the $s$-triazine sheet. (b). The dependence of strain energy (Ry) on uni-axial tensile strain of the $s$-triazine sheet. (c). The dependence of strain energy (Ry) on area ( $\mathit{Angstrom}^2$ of the $s$-traizine sheet for bulk modulus estimation.
}
\end{center}
\end{figure}

\begin{figure}
\label{fig3}
(a) \includegraphics[width=3.2083in,height=2.2291in]{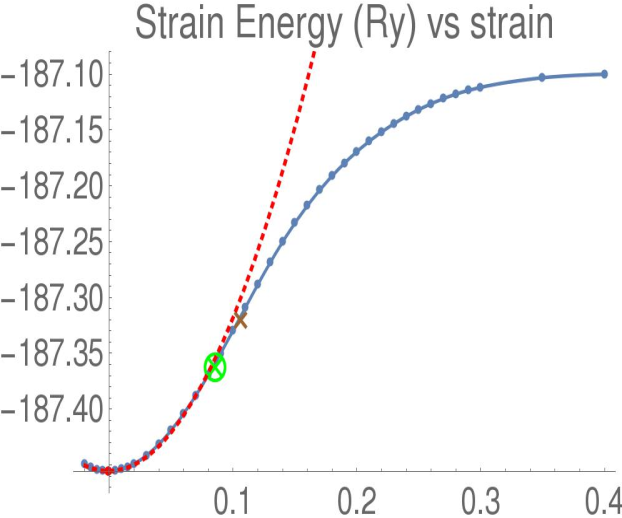}
(b) \includegraphics[width=3.2083in,height=2.2291in]{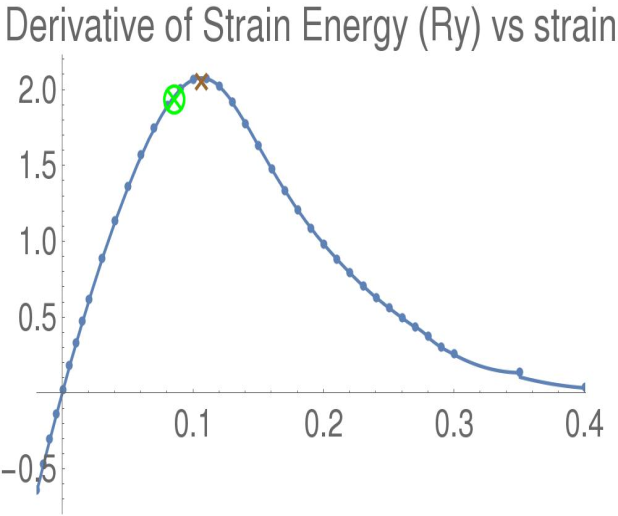}
(c)  
\includegraphics[width=3.2083in,height=2.2291in]{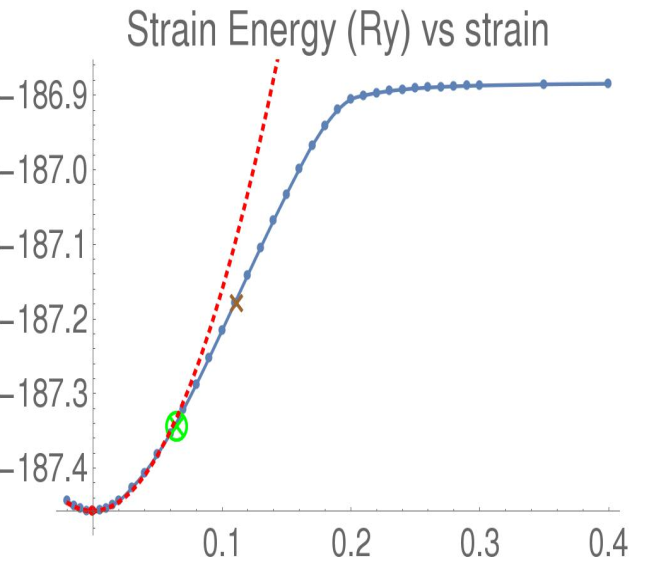}
(d)  
\includegraphics[width=3.2083in,height=2.2291in]{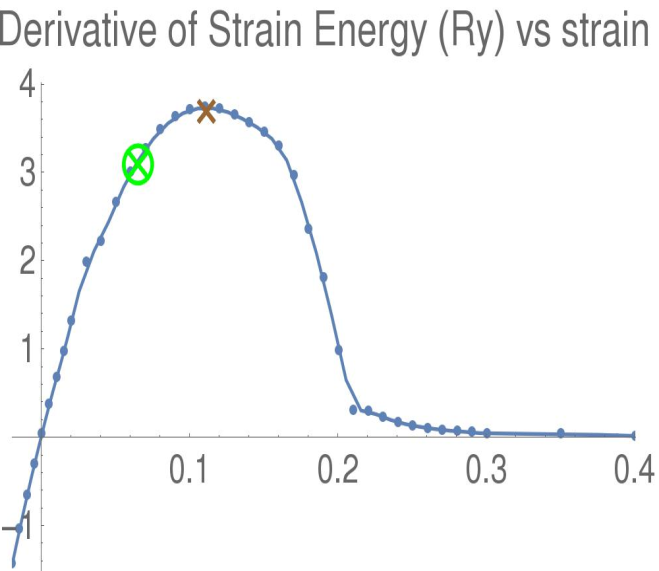}
\caption{
Dependencies of strain energy and derivative of energy on strain of $s$-triazine sheet. (a), (b) Uni-axial strain; (c), (d) bi-axial strain. The two critical points are labeled as $s_{\mathrm m1}$ (circled star) and $s_{\mathrm m2}$ (star) in the Figures. The insets in 3a and 3c are strain energy curves in harmonic elastic region. The dots are raw DFT data points computed in this work, while the continuous lines are best-fit curves to these data points. The dotted curves are harmonic potentials fitted to the data points centered in the  $\pm 2\mathit{}$ strain region.
}
\end{figure}

\begin{figure}
\begin{center}
\label{fig4}
\includegraphics[width=3.2083in,height=2.2291in]{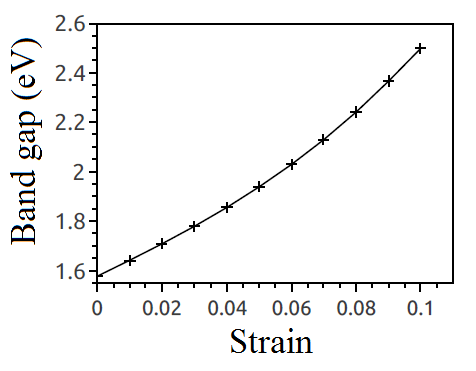}
\caption{The dependence of bandgap on uniform bi-axial tensile strain for $s$-triazine sheet.}
\end{center}
\end{figure}

\begin{figure*}
\begin{center}

\label{fig5}
(a)  
\includegraphics[width=1.89in,height=2.04in]{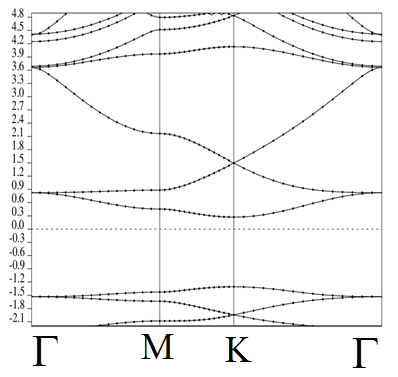}
(b)  
\includegraphics[width=1.89in,height=2.04in]
{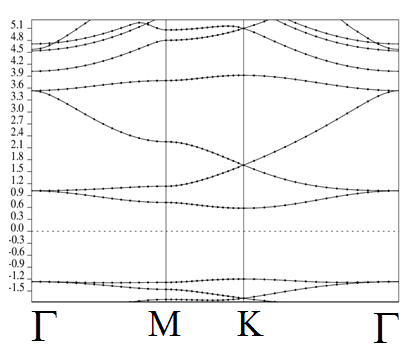}
 
(c)  
\includegraphics[width=2.2in,height=2.04in]{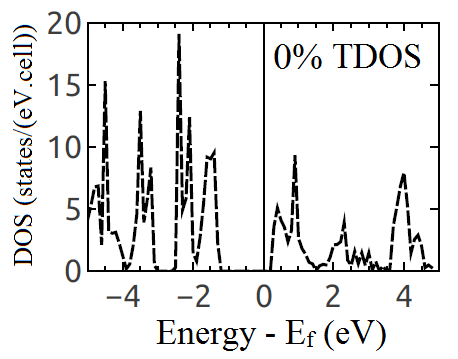}
(d)  
\includegraphics[width=2.2in,height=2.04in]
{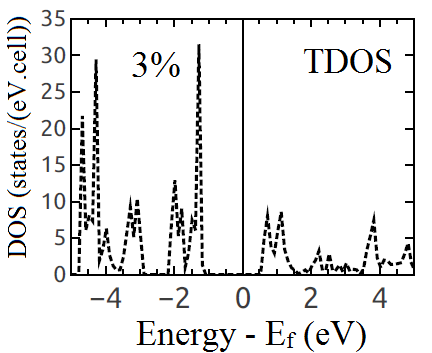}

(e)  
\includegraphics[width=2.2in,height=2.04in]{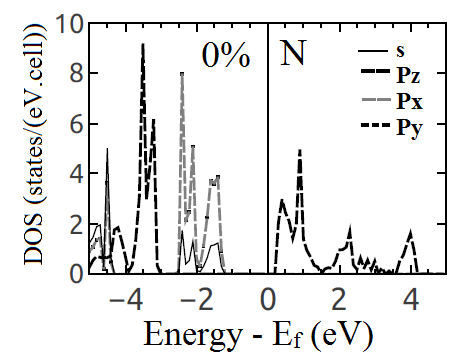}
(f) 
\includegraphics[width=2.2in,height=2.04in]{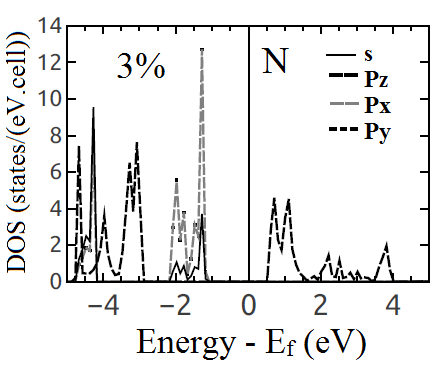}

(g)  
\includegraphics[width=2.2in,height=2.04in]
{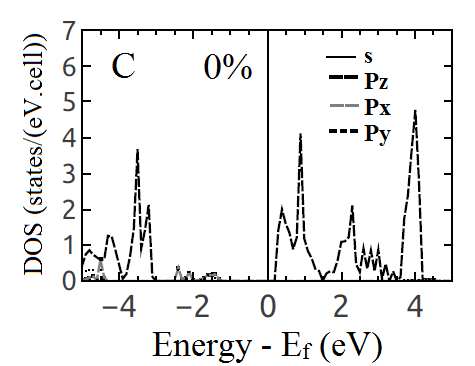}
(h)  
\includegraphics[width=2.2in,height=2.04in]
{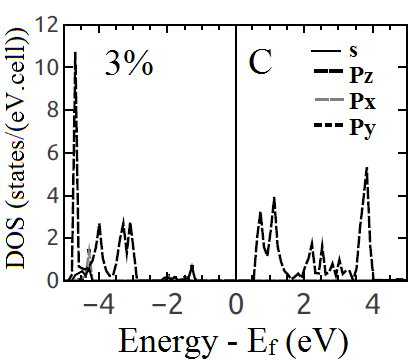}
\end{center}
\end{figure*} 

\begin{figure}
\begin{center}
(i)  
\includegraphics[width=2.2in,height=2.04in]
{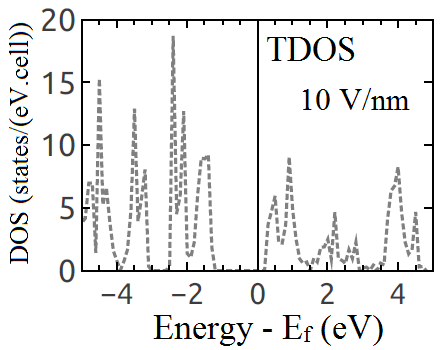}
\caption{
The band structure, and the corresponding total and projected densities of state for the strained ($s$ = 3\%) and unstrained ($s$ = 0) $s$-triazine systems.
(a) and (b) Band structures for the $s$-triazine systems respectively. (c) and (d) The total density of states (TDOS) for the $s$-triazine system.
(e)-(h) The projected density of states (PDOS) for sp-like orbitals of the sum of N and C atoms in the $s$-triazine systems respectively.
(i) The TDOS of the $s$-triazine sheet under electric field of magnitude 10.0 V/nm.
}
\end{center}
\end{figure}

\end{document}